\documentclass{article}
\usepackage{amstex,amssymb,psfig}
\begin{document}

\title{Josephson Junctions with Minimal Length}
\author{T. L. Boyadjiev\\
{\small \textit{Faculty of Mathematics and Computer Science,}}\\
[-1.mm]{\small \textit{Sofia University 'St. Kliment Ohridski', Bulgaria}}\\
[-1.mm]{\small \textit{e-mail: todorlb@@fmi-uni.sofia.bg}} \\
\\
M. D. Todorov\\
{\small \textit{Faculty of Applied Mathematics and Computer Science,}}\\
[-1.mm]{\small \textit{Technical University of Sofia, Bulgaria}}\\
[-1.mm]{\small \textit{e-mail: mtod@@vmei.acad.bg}}}
\date{}
\maketitle

\bigskip

\hspace{1cm}{\bf PACS:} 85.25.Cp, 85.25.Am, 02.60.Lj

\bigskip

\begin{abstract}
The minimal length of ``one-dimensional'' Josephson junctions, in
which the specific bound states of the magnetic flux retain their
stability is discussed numerically. Thereby, we consider as
``long'' every Josephson junction, in which there exists at least
one nontrivial stable distribution of the magnetic flux for fixed
values of all the physical and the geometrical parameters.

Our results can be applied for optimization of the sizes of
devices containing JJs for different operating conditions.
\end{abstract}

\section{Formulation of the Problem}

It is known that the stationary distributions (bound states) of
the magnetic flux $\varphi (x)$ in  ``long'' (one-dimensional)
Josephson junctions (JJs) can be modeled as solutions of the
nonlinear boundary value problem (BVP):
\begin{equation}
-\varphi_{xx}+j_{D}(x)\sin \varphi +\gamma=0, \quad \varphi_{x}
({\pm R})=h_{B}, \label{sg}
\end{equation}
\noindent where $\gamma$ is the bias current, and $h_{B}$ is the
external (boundary) magnetic field. We note that the kind of
boundary conditions, either in presence or in absence of the
current $\gamma $ in the right-hand side of eqn. (\ref{sg}), is
determined by the geometry of the junction. Here we consider
simple JJs with overlap geometry \cite{barone}, in which the
current $\gamma $ can be approximately considered as a constant.
The generalization of our results in the case of another
geometry, for example in-line geometry, does not require big
efforts.

We suppose that the given function $0\le j_{D}(x)\le 1$ describes
the variations of the Josephson current amplitude, caused by the
possible local inhomogeneities of the dielectric layer thickness
\cite{and62}. When $j_{D}(x)\equiv 1$ the junction is homogeneous
(see the complete analytical results in recent paper
\cite{cfgsv00}).

The single inhomogeneity is characterized by its respective
position $x_0$ and width $\mu$, so the amplitude $j_{D}= j_{D}(x,
x_{0}, \mu)$. The frequently used expression  $j_{D}(x) = 1- \mu
\delta(x-x_0)$ (here $\delta(x)$ is Dirac's function) is useful
in theoretical examination (see the basic papers  \cite{mkls78}
and \cite {galpfil}). In this paper we use an isosceles trapezium
with base $\mu $ (Fig.1) as a more suitable physical model of
inhomogeneity \cite{alexboyad, tedmi00}.

\vspace{0.5cm} \centerline{\psfig{figure=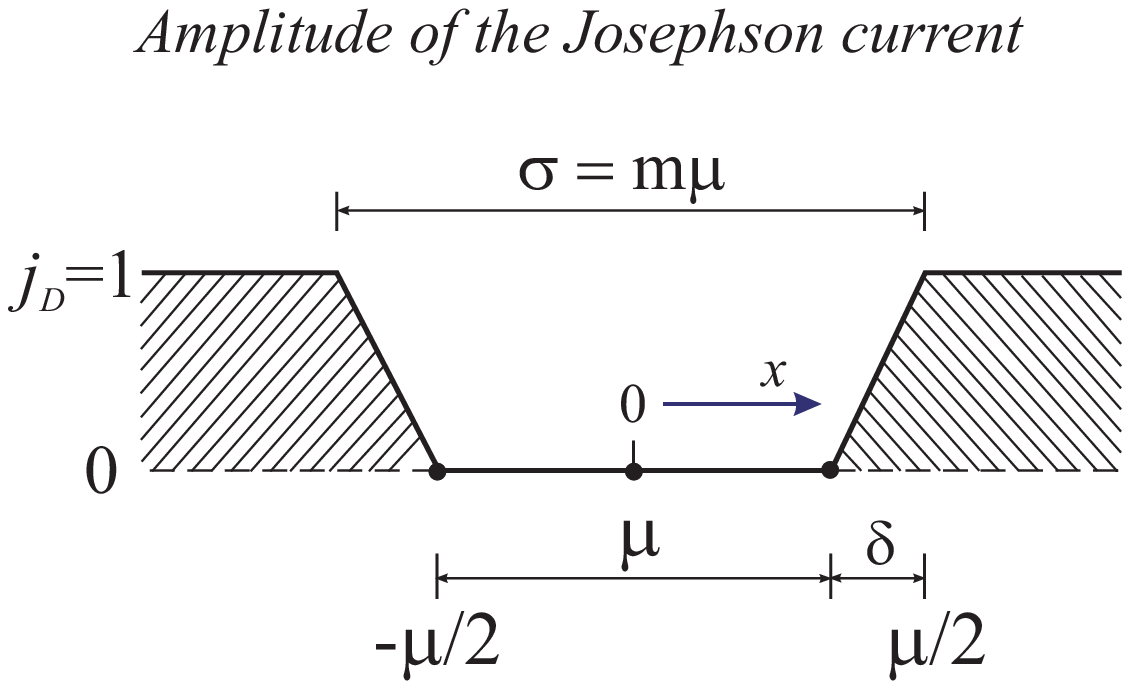,width=4.in}}
\centerline{\small Figure 1} \vspace{0.5cm}

The stability of some concrete solution $\varphi (x)$ of the BVP
(\ref{sg}) is determined by the sign of the minimal eigenvalue
$\lambda_{min}$ of the corresponding Sturm-Liouville problem
(SLP):
\begin{equation}
-\psi_{xx}+q(x)\;\psi =\lambda \,\psi \>,\quad \psi_{x}(\pm R)=0\>,  \label{slp}
\end{equation}
\noindent where $q(x)=j_{D}(x)\cos \varphi (x)$ is the potential
of SLP, originated from the mentioned solution $\varphi (x)$.

Each of the different eigenvalues $\lambda_{min}\equiv
\lambda_{0}<\lambda_{1}<\lambda_{2}<\; \dots \;
\lambda_{n}<\;\dots \;,$ conforms to an unique eigenfunction
$\psi_{n}(x)\>,$ $\>n=0,1,2,\dots \;$, which satisfies some
additional (norm) condition, for example:
\begin{equation}
\int\limits_{-R}^{R}\psi ^{2}(x)\;d\,x=1\>. \label{norm}
\end{equation}

If the minimal eigenvalue $\lambda_{min}>0\>$, then the
respective solution $\varphi (x)$ of BVP (\ref{sg}) is stable
with respect to small time-space perturbations. If the minimal
eigenvalue $\lambda_{min}<0 $, then this solution is unstable.
The value $\lambda_{min}=0$ corresponds to a bifurcation point, in which the
stable solutions of eqn. (\ref{sg}) go to unstable ones and vice
versa (for details see \cite{galpfil}).

Apart from the space coordinate $x$, the virtual solutions of the
nonlinear BVP (\ref{sg}) depend also on the physical parameters
$h_{B}$, $\gamma $ and the ``technological'' ones $\mu $, $\Delta
= 2R$ (length of the junction), and $x_0$, i.e., $\varphi
=\varphi(x,p)$, where we simply substitute $p\equiv \{h_{B},
\gamma, \mu, x_0,  \Delta \} $. The varying of each of those
parameters causes a variation of the distribution $\varphi (x,p)$
and, therefore, subsequent variations of the potential $q(x,p)$,
the eigenvalues $\lambda (p)$, and the respective eigenfunctions
$\psi (x,p)$. Thus, we can conclude that every concrete solution
of BVP (\ref{sg}) has an area, where this solution remains stable
with regard to the variations of the parameters $p$. In the
parametric space the equation:
\begin{equation}
\lambda_{min}(p)=0 \label{eq:7}
\end{equation}
\noindent determines a hypersurface, whose points appear to be
bifurcation points corresponding to the solution under
consideration. The intersections of the bifurcation hypersurface
(\ref{eq:7}), when there are chosen pairs of the parameters $p$
for fixed values of others, are called bifurcation curves and the
respective values of the parameters - bifurcation (critical)
parameters. From a physical viewpoint most interesting seem to be
the bifurcation curves of the kind ``current - magnetic flux''
$\lambda_{min} (h_{B}, \gamma )=0$, when the geometrical
parameters $\mu $, $R$, and $x_0$ are fixed. These curves could
be obtained experimentally \cite{vystavkin, lmu94, mu94}.

\vspace{0.5cm} \centerline{\psfig{figure=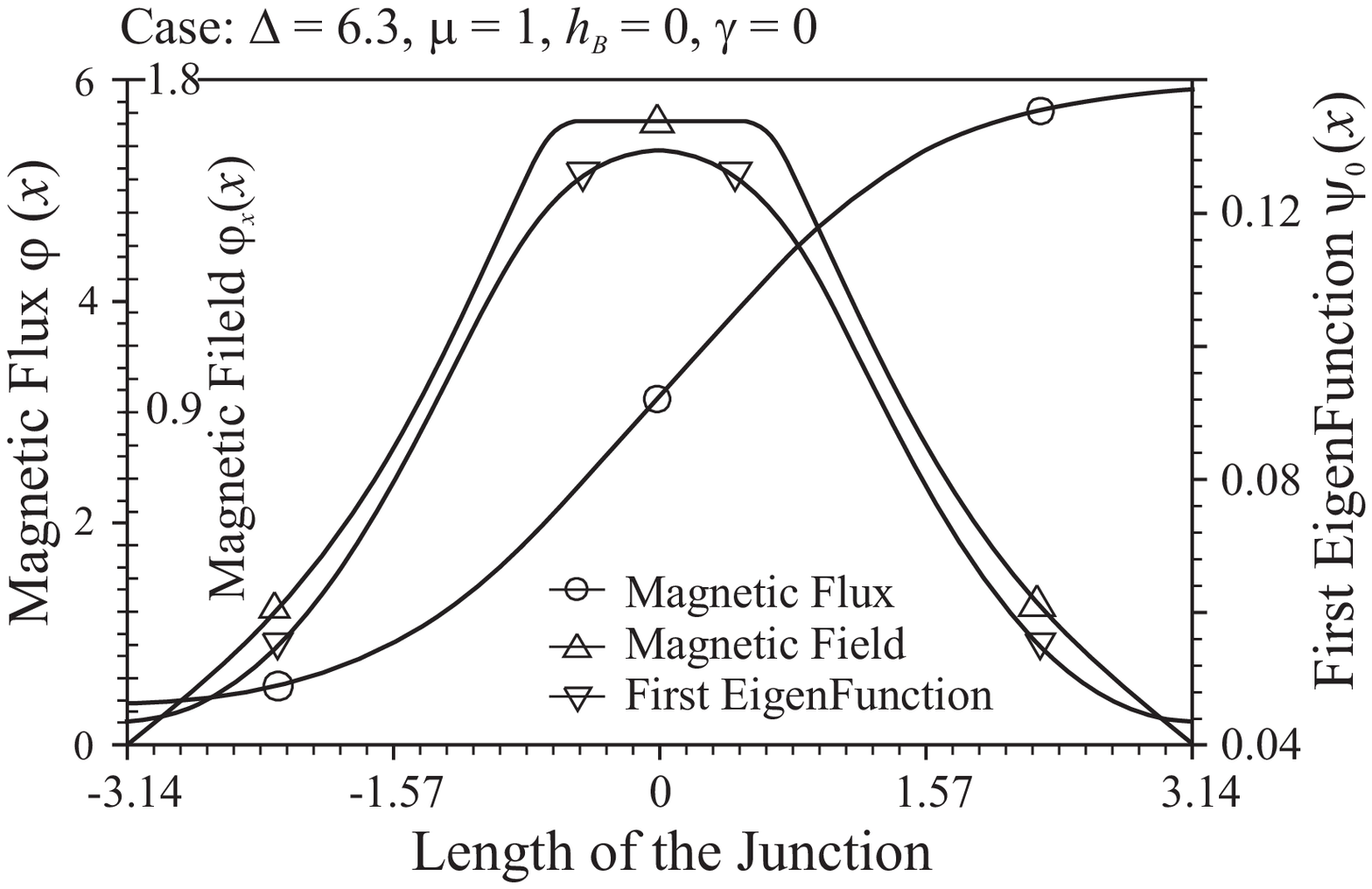,width=4.in}}
\centerline{\small Figure 2} \vspace{0.5cm}

From a technological point of view, however, it is worth
investigating the bifurcation curves as functions with respect to
at least one of the geometrical parameters, for example:
\[
\lambda_{min}(h_{B},\Delta)=0\>, \lambda_{min}(\gamma ,\Delta)=0\>,\>
\lambda_{min}(\mu ,\Delta)=0,\> \dots
\]
\noindent Such kind of problems can be connected with the
optimization of sizes of devices, containing JJs.

Most of our results below are related to the typical solution of
BVP (\ref{sg}), physically existing both in homogeneous and
inhomogeneous JJs: so called single fluxon. Let us note that in
the infinite homogeneous or inhomogeneous JJ with one
$\delta$-shaped microinhomogeneity at the point $x_0=0$, the
single fluxon/antifluxon is represented by the exact solution
\cite{galpfil}, \cite{barone}:
\begin{equation}\label{single}
 \varphi (x)=4\arctan \exp (\pm x).
\end{equation}

More complicated (multi-fluxon) solutions there exist in
inhomogeneous JJs for big enough values of $h_B$.

The numerically obtained basic functions: the magnetic flux
$\varphi(x)$, the magnetic field $\varphi_x(x)$, and the first
eigenfunction $\psi_0(x)$ of such solution for $\Delta = 6.3$,
$\mu = 1$, $h_B = 0$, and $\gamma = 0$ are represented at Fig. 2.

Fig. 3 represents the typical relationships $\lambda_{min}(\Delta
)$, corresponding to the single fluxons in inhomogeneous
junctions with one resistive inhomogeneity placed at the point
$x_0=0$ when $\mu =0.5$ and $\mu =1$, respectively. The points
$B_{0}$ and $B_{1}$ appear to be bifurcation points for these
solutions, i.e., the zeroes of the corresponding relationships
$\lambda_{min}(\Delta )$. According to the above mentioned
reasonings, these points determine the minimal length of the JJ,
for which the single fluxon is still stable for given values of
the other parameters.

\vspace{0.5cm}
\centerline{\psfig{figure=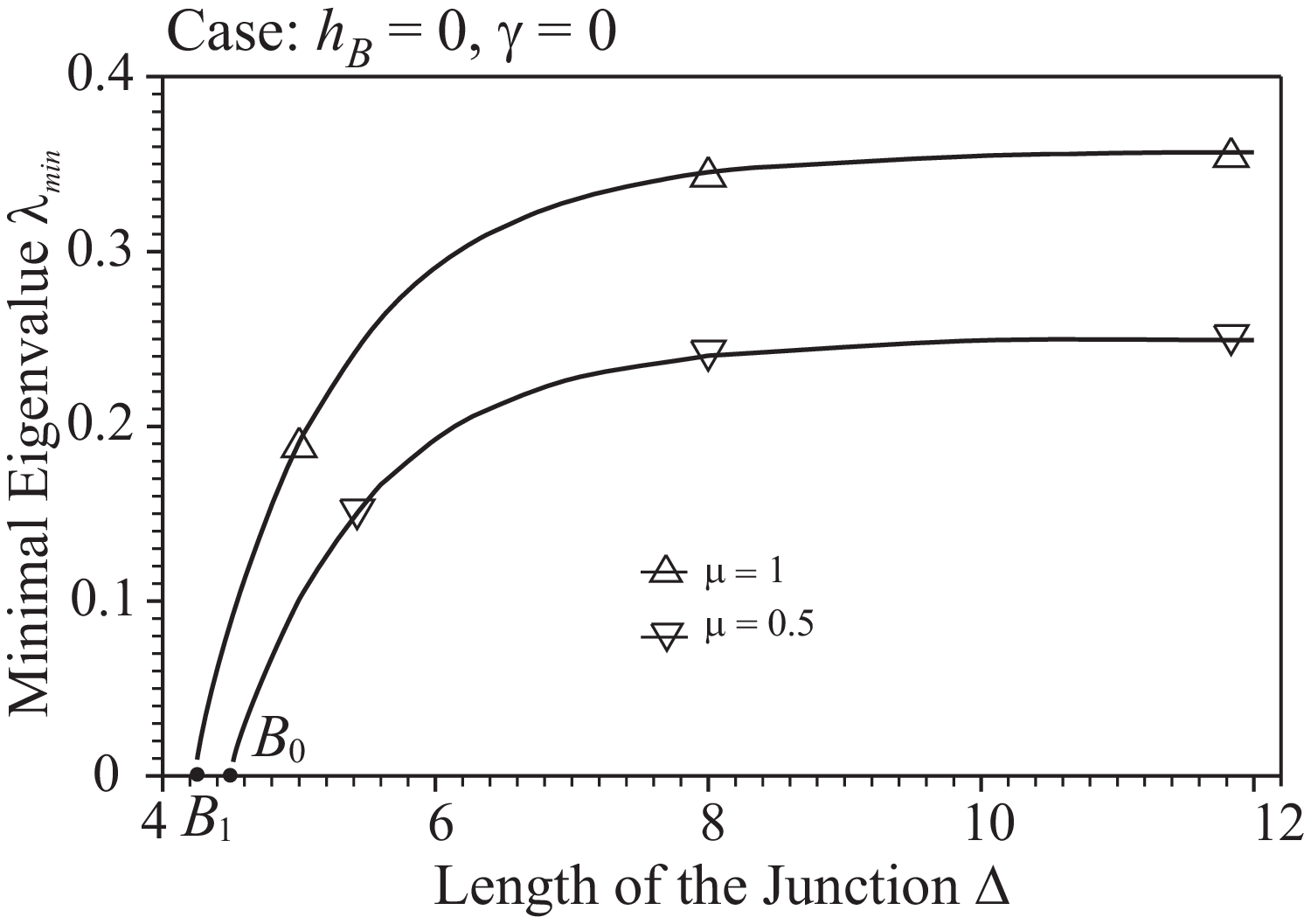,width=4.in}}
\centerline{\small Figure 3} \vspace{0.5cm}

Usually, from physical point of view, the concrete JJ is ``long"
or ``one-dimensional" if the Josephson penetration depth
$\lambda_J$ satisfies the condition $\lambda_J >> 1$. However,
such definition is not perfect enough mathematically. On the
other hand, it seems that the calculation of the minimal
geometrical sizes and, especially, the minimal length $\Delta$ of
homogeneous or inhomogeneous JJs, corresponding to a concrete
nontrivial distribution of the magnetic flux, is an important
practical problem.

The trivial method, obviously, is through the point by point
construction of the curves of type $\lambda_{min}(\Delta)$ using
BVPs (\ref{sg})-(\ref{norm}). However, it is quite natural to put
the question how to calculate directly the bifurcation curves. A
general approach of study for the formulated problem is proposed
in the papers \cite {boyad87, boyad88}.

We consider the system of the equations (\ref{sg}), (\ref {slp}),
and (\ref{norm}) as a closed nonlinear eigenvalue problem with
respect to the functions $\varphi (x)$, $\psi (x)$ and one of the
parameters $p$, for example, $h_{B}$ or $\gamma $ in
\cite{boyad88}, or $R$ in this case, while the other parameters,
and the eigenvalue $\lambda $ as well, are given. Fixing $\lambda
>0$ to be small enough (for example $\lambda =0.001 \div 0.01$),
we guarantee a stable solution to be obtained. We note, that the
derivative ${\partial \,\lambda }/{\partial \,p_{0}}\to \infty $
when $p_{0}$ approaches its critical (bifurcation) value, so, the
solutions of the above system with a priori prescribed accuracy
belong to the small vicinity of the sought bifurcation curve.

The generalized Continuous Analog of Newton's Method \cite{paz}
for solving the mentioned system is used (see details in our
recent paper \cite {tedmi00}).

We have to note that for some initial approximations the method
can lead to an unstable solution, as well. In this case, the
given $\lambda$ corresponds to one of the higher eigenvalues, not
to the minimal one. The simplest method to avoid this difficulty
uses the oscillation theorem \cite{levitan} for eigenfunctions:
{\it If the obtained function $\psi(x)$ is the minimal
eigenfunction, then it does not have a zero on the interval
$\Delta$}.

\section{Results and Discussion}

Fig. 4 shows the convergence of the iteration process. The upper
curve, marked by ``$\nabla $'', presents the initial distribution
of the magnetic field $\varphi_{x}(x)$ alongside the junction.
This distribution is chosen to be a solution of BVP (\ref{sg}),
for $\Delta=10$, $\mu = 1$, $h_{B}=0$, and $\gamma =0$. At the
same figure the lower curve, marked by ``o'', presents the
calculated distribution of the magnetic field corresponding to
the minimal eigenvalue $\lambda_{min}=0.01$. Hence, this is the
sought bifurcation distribution and, in the framework of this
model, the value of the spectral parameter $\Delta_{min}\approx
4.24$ is the minimal length of the junction, providing a stable
single fluxon. In this sense, junctions, whose length lies below
the critical value $\Delta_{min}$ of $\Delta$, have to be
considered as ``short'' for distributions $\varphi(x)$ of the
kind single fluxon. However, for such a short length $\Delta$
there are a unique stable and unstable distributions of the
magnetic flux in the junction - Meissner's solutions: the trivial
solutions $\varphi (x)=0, 2\pi, \dots$ (stable) and $\varphi
(x)=\pi, 3\pi, \dots $ (unstable) for $h_{B}=0$ and $\gamma =0$.

\vspace{0.5cm} \centerline{\psfig{figure=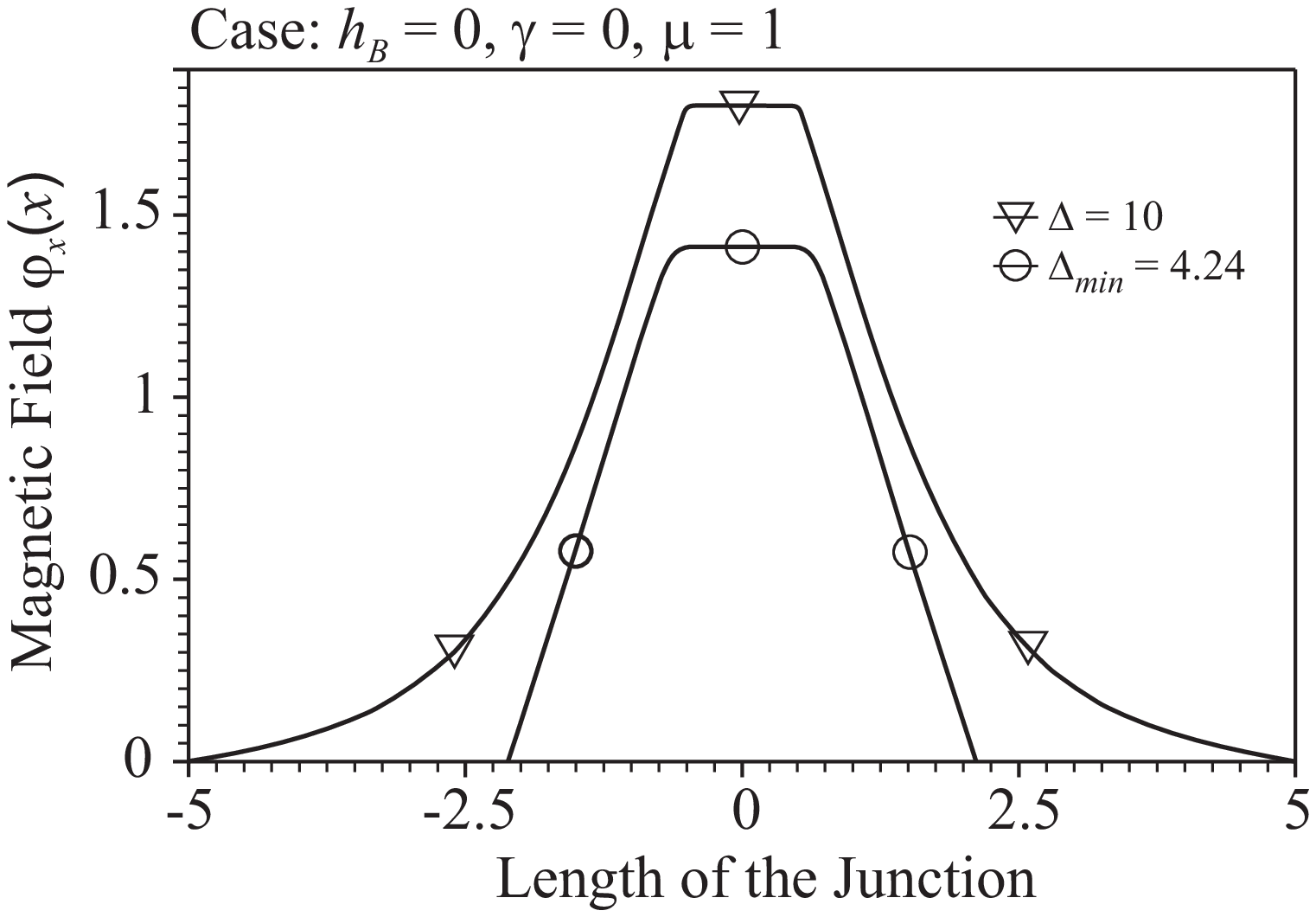,width=4.in}}
\centerline{\small Figure 4} \vspace{0.5cm}

The influence of the parameters $h_B$, $\gamma$, $\mu$, and $x_0$
on the minimal length $\Delta_{min}$ of the JJs will be
considered below.

If the current $\gamma \neq 0$ and the values of all other
parameters are  fixed, then the stability of the fluxon requires
the junction to be longer (see Fig. 5). This fact corresponds to
the classical conclusion \cite{mkls78} that the existence of
quantity $\gamma$ violates the stability of the bound states in
JJs. The ``force" $\gamma$ displaces the maximum of
$\varphi_x(x)$ from $x_0=0$ to the left or right depending on its
sign; the full magnetic flux $\Delta \varphi = \varphi(R)
-\varphi(-R)$ increases, so that the minimal junction's length
$\Delta_{min}$, necessary to accommodate such a flux, increases
as well.

\vspace{0.5cm} \centerline{\psfig{figure=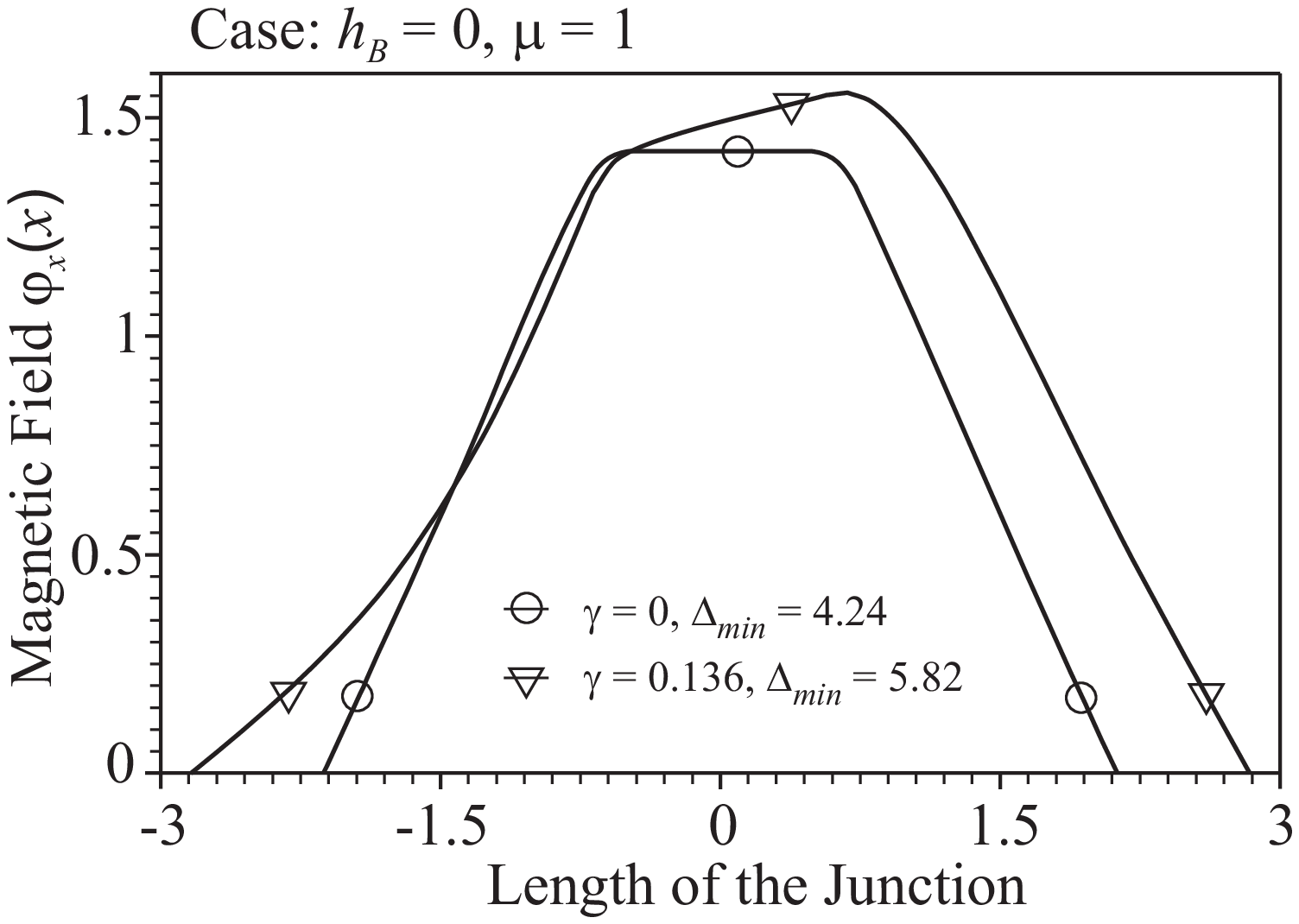,width=4.in}}
\centerline{\small Figure 5} \vspace{0.5cm}
\vspace{0.5cm}
\centerline{\psfig{figure=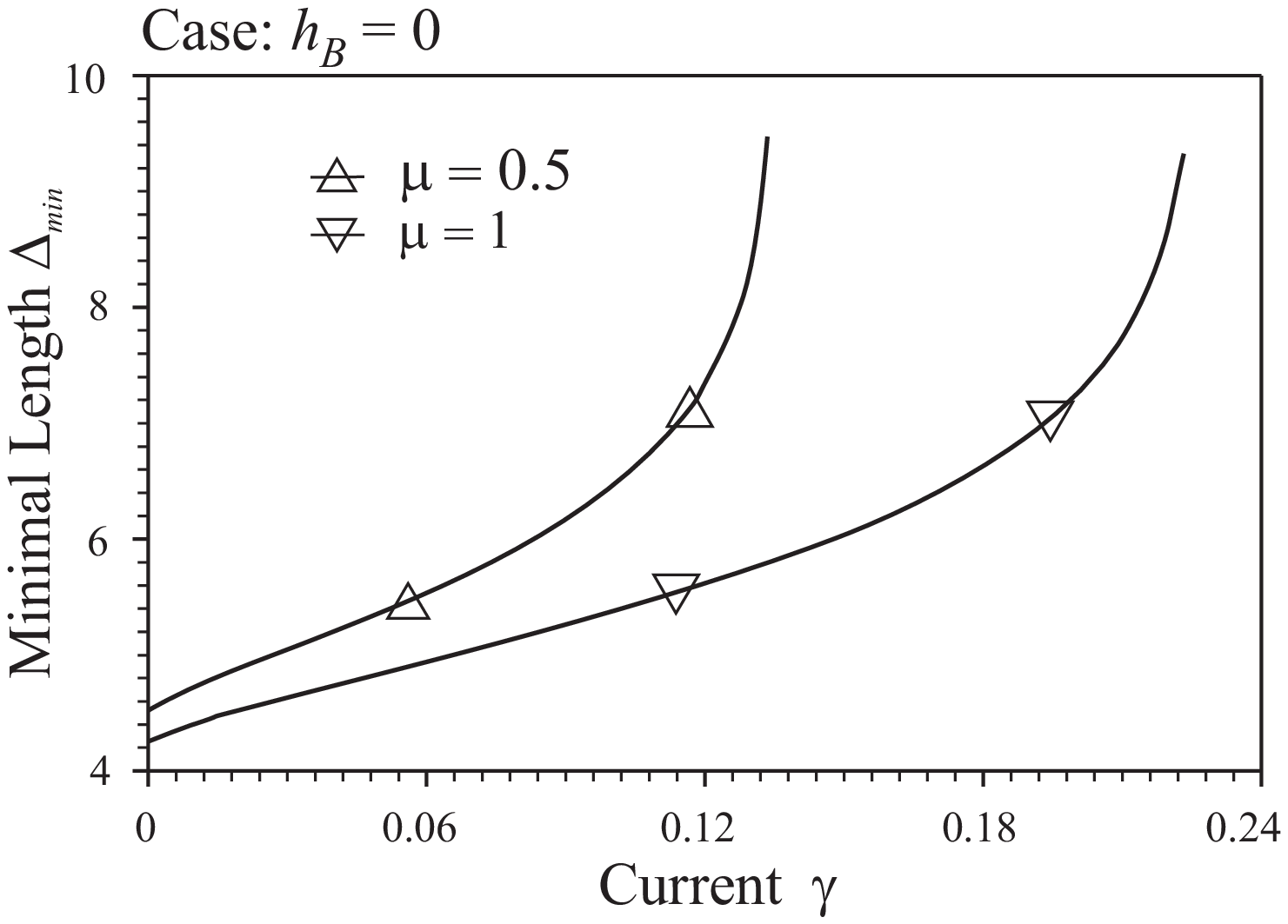,width=4.in}}
\centerline{\small Figure 6} \vspace{0.5cm}

The numerically obtained influence of the current $\gamma$ upon
the bifurcation length $\Delta_{min}$ is plotted on Fig.6 for two
values of the size $\mu$ of the inhomogeneity: $\mu =0.5$ and
$\mu = 1$. The curves increase very abruptly and when $\gamma$
approaches some critical value, the derivative ${\partial
\Delta_{min}}/{\partial \gamma} \rightarrow \infty$. This
behaviour means that beyond the critical value of $\gamma$, JJs
which correspond to the single fluxon do not exist.

\vspace{0.5cm} \centerline{\psfig{figure=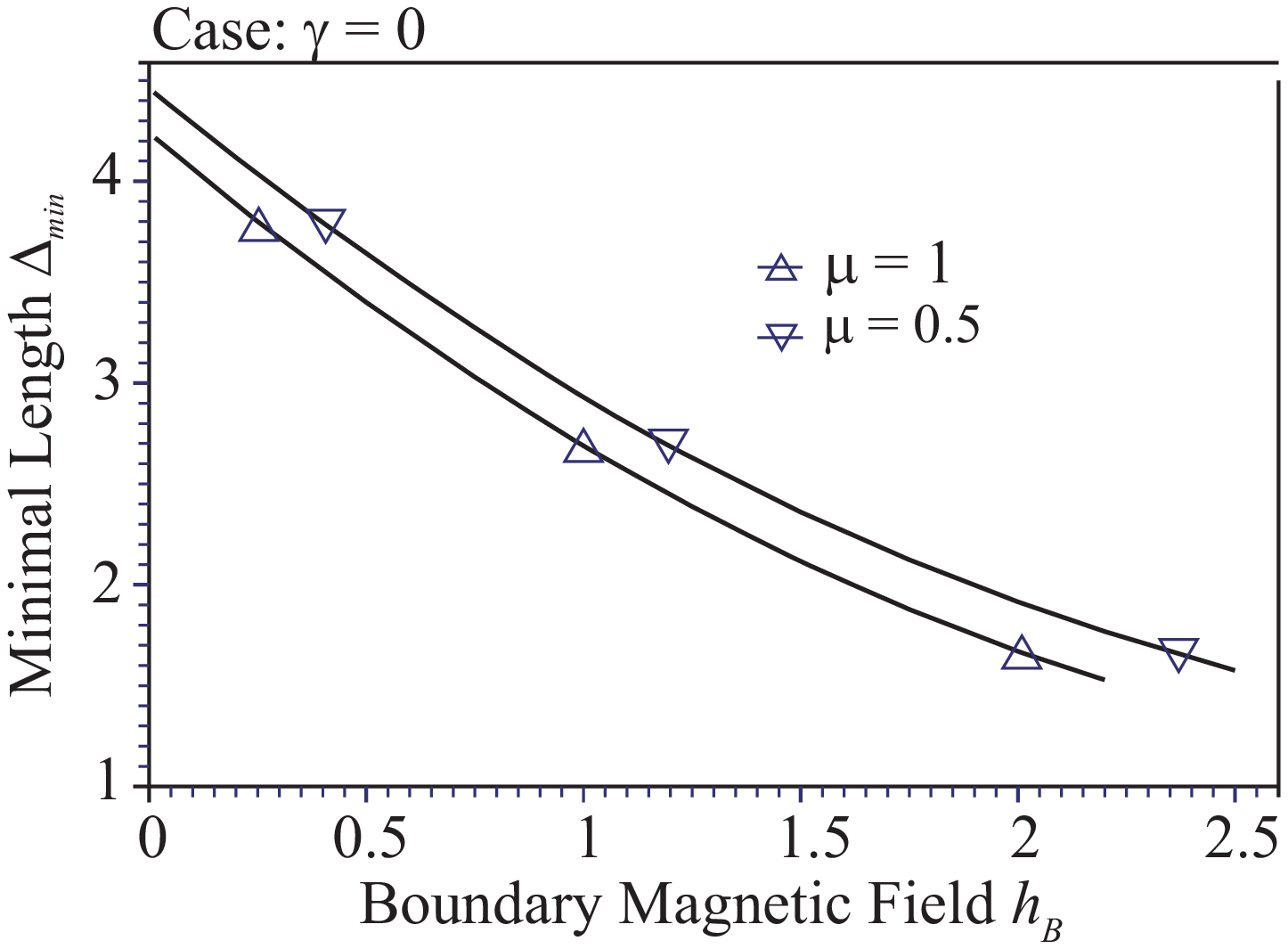,width=4.in}}
\centerline{\small Figure 7} \vspace{0.5cm}

On the contrary, if the boundary magnetic field $h_B \neq 0$ and
varies within some fixed domain, then the stability of the fluxon
is improved (see Fig. 7). Well noticeable is the stabilizing
influence of the boundary magnetic field $h_B$ upon the critical
length of the junction, i.e., the increase of the magnitude of
$h_B$ decreases the minimal length of the junction
$\Delta_{min}$. Thus, the calculated minimal length of the JJ
corresponding to the single fluxon ($h_{B}=0,$ $\gamma =0$) is
$\Delta_{min} \approx 4.24$, while the same quantity when
$h_{B}=1$ and $\gamma =0$ is $\Delta_{min}\simeq 2.7$.

\vspace{0.5cm} \centerline{\psfig{figure=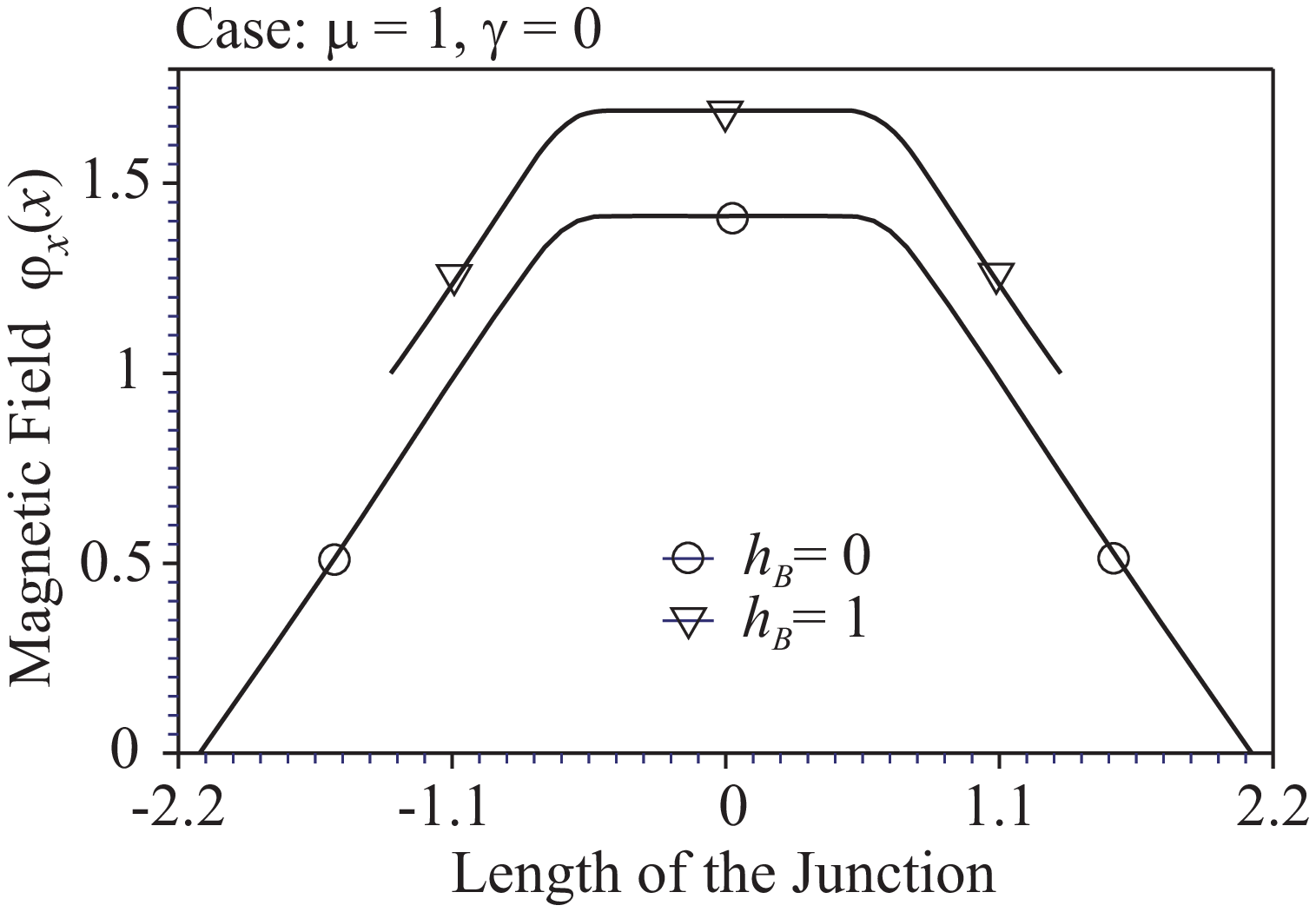,width=4.in}}
\centerline{\small Figure 8} \vspace{0.5cm}

Fig.8 represents the distributions of magnetic field
$\varphi_{x}(x)$ alongside the junction in absence of current
($\gamma =0$) depending on its boundary values $h_{B}=0$ and
$h_{B}=1$, respectively. It is clear that the two solutions could
be well approximated by means of two-degree polynomials.
Furthermore, they seem to be geometrically similar. Namely, the
curve $\varphi_{x}(x)$ corresponding to a magnetic field
$h_{B}>0$ may be considered as obtained from the other curve by a
homothety alongside $x$ axis with respect to the pole $x=0$ and a
translation alongside the axis $\varphi_{x}$.

\vspace{0.5cm} \centerline{\psfig{figure=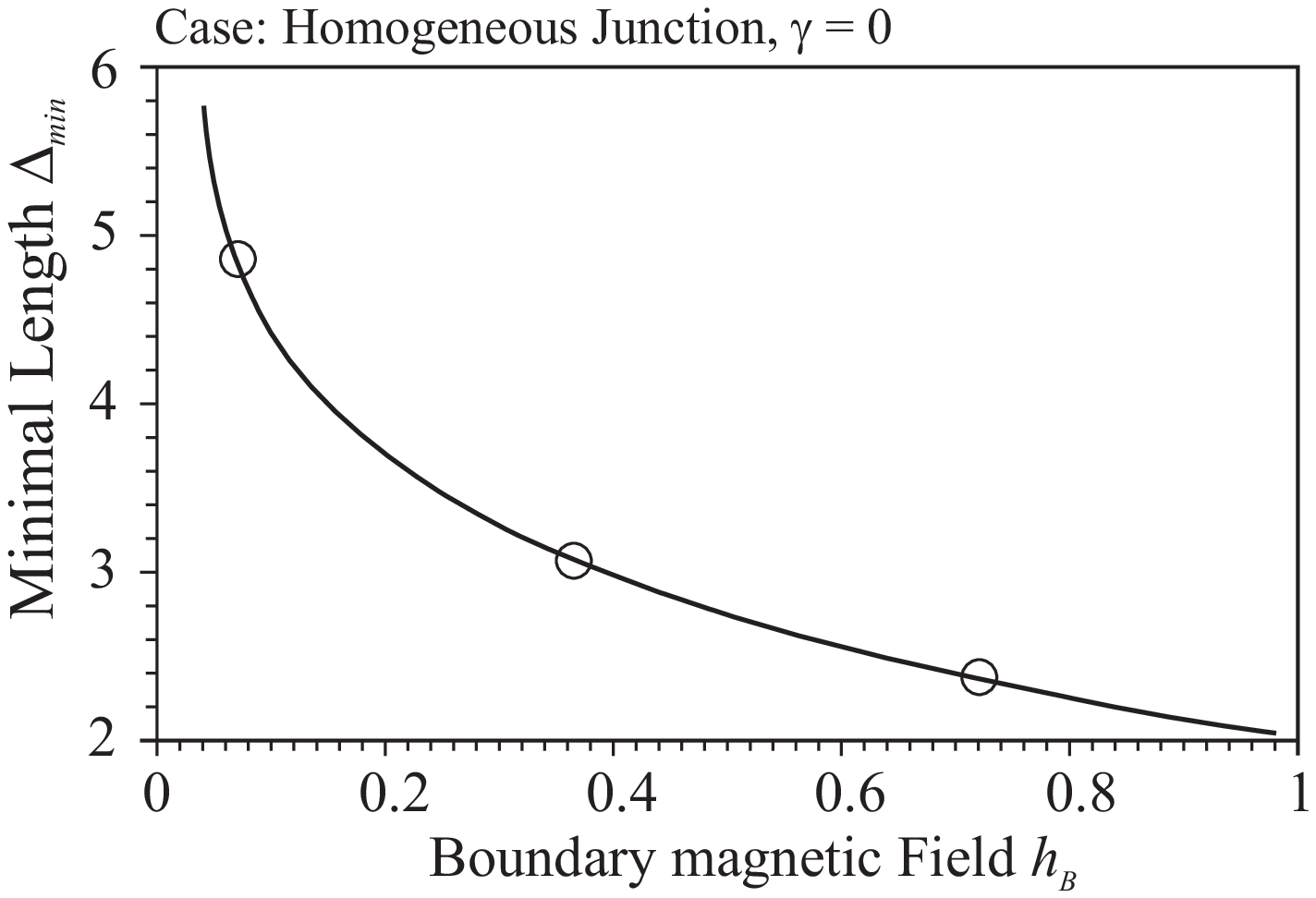,width=4.in}}
\centerline{\small Figure 9} \vspace{0.5cm}

At the next Fig.9, the dependence of the minimal length
$\Delta_{min}$ on the boundary magnetic field $h_B$ for $\gamma =
0$ in the case of a homogeneous JJ (the amplitude $j_D(x) \equiv
1$) is presented. Even though this dependence is qualitatively
similar with the analogous ones at Fig.7, it differs from the big
slope of $\Delta_{min}$ for small $h_B$, so that
$\Delta_{min}(h_B) \rightarrow \infty$ when $h_B \rightarrow 0$.

Such behaviour of the curve $\Delta_{min}(h_B)$ can be
qualitatively explained by means of the formula \cite {galpfil}:
\begin{equation}
\lambda_0=\lambda_{min} = \frac{\mu}{2} \left[ \sqrt {1+
\frac{\mu^2}{16}}- \frac{\mu}{4} \right], \label {theor}
\end{equation}
\noindent which is exact for the single fluxon (\ref{single}) in
an infinite JJ with one resistive $\delta$-shaped inhomogeneity
at the center $x=0$. Hence, for $\mu = 0$ we obtain
$\lambda_{min} = 0$, so that for $h_B = 0$ and $\gamma=0$ the
single fluxon in homogeneous JJ has to be considered as a
quasi-stable (bifurcation) state of the magnetic flux
$\varphi(x)$ \cite{galpfil}. On the contrary, in inhomogeneous JJ
with resistive inhomogeneities the single fluxon is still stable
($\lambda_{min} > 0$), which physically is due to the existence
of the inhomogeneity.

In the case of an inhomogeneous JJ with finite length
$\Delta=6.3$ the calculated curve $\lambda_{min}(\mu)$ (marked by
``$\square$") for $h_B = 0$ and $\gamma = 0$ is shown at Fig.10.
We note that for small enough values of the size $\mu$ the
dependence is analogous to (\ref{theor}).

Let us introduce the full energy $F$ connected with some solution
$\varphi(x)$ of BVP (\ref{sg}) via the formula:
$$F(\mu)=\int\limits_{-R}^R \left[ \frac{1}{2} \varphi_x^2 + j_D(x)
\left( 1-\cos \varphi \right) + \gamma \varphi \right] dx + h_B
\left[ \varphi(R) - \varphi(-R) \right].$$ \noindent It is
efficient to compare the energy curves $F(\mu)$ (marked by
``$\triangledown$") for fixed $\Delta=6.3$ and $F_{min}(\mu)$
(marked by ``$\circ$") for a JJ with minimal length (see Fig.
10). The magnitudes of the energy are regarded to the energy
($F=8$) of the single fluxon (\ref{single}) in an infinite
homogeneous JJ. It is well noticeable that for $\mu > 0$ the
inequality $F(\mu) > F_{min}(\mu)$ is satisfied, e.g., the single
fluxon in the JJ with minimal length is more stable. The cusp for
$\mu = 0$ corresponds to a bufurcation in accordance with formula
(\ref{theor}).

\vspace{0.5cm} \centerline{\psfig{figure=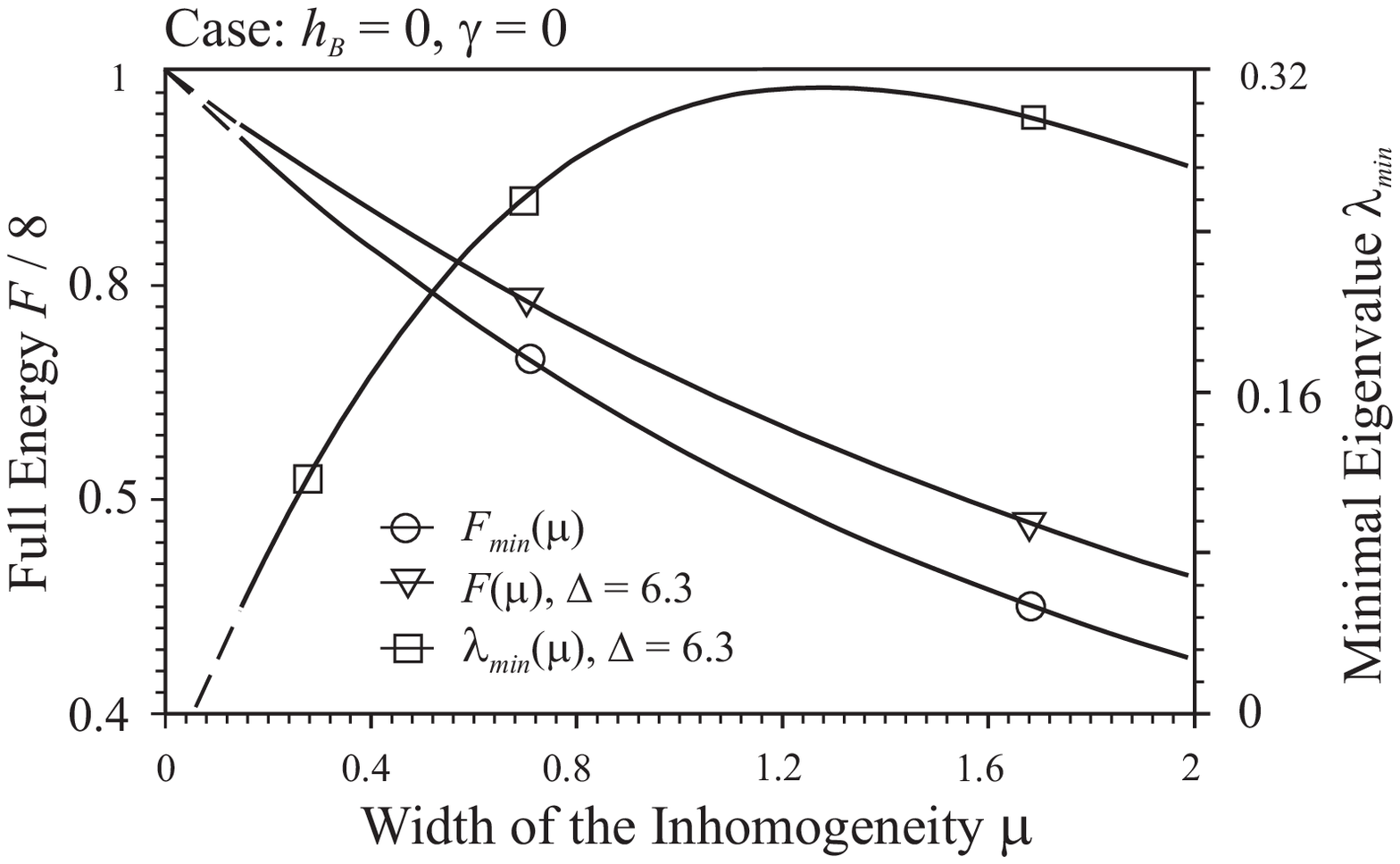,width=4.in}}
\centerline{\small Figure 10} \vspace{0.5cm}
\vspace{0.5cm}
\centerline{\psfig{figure=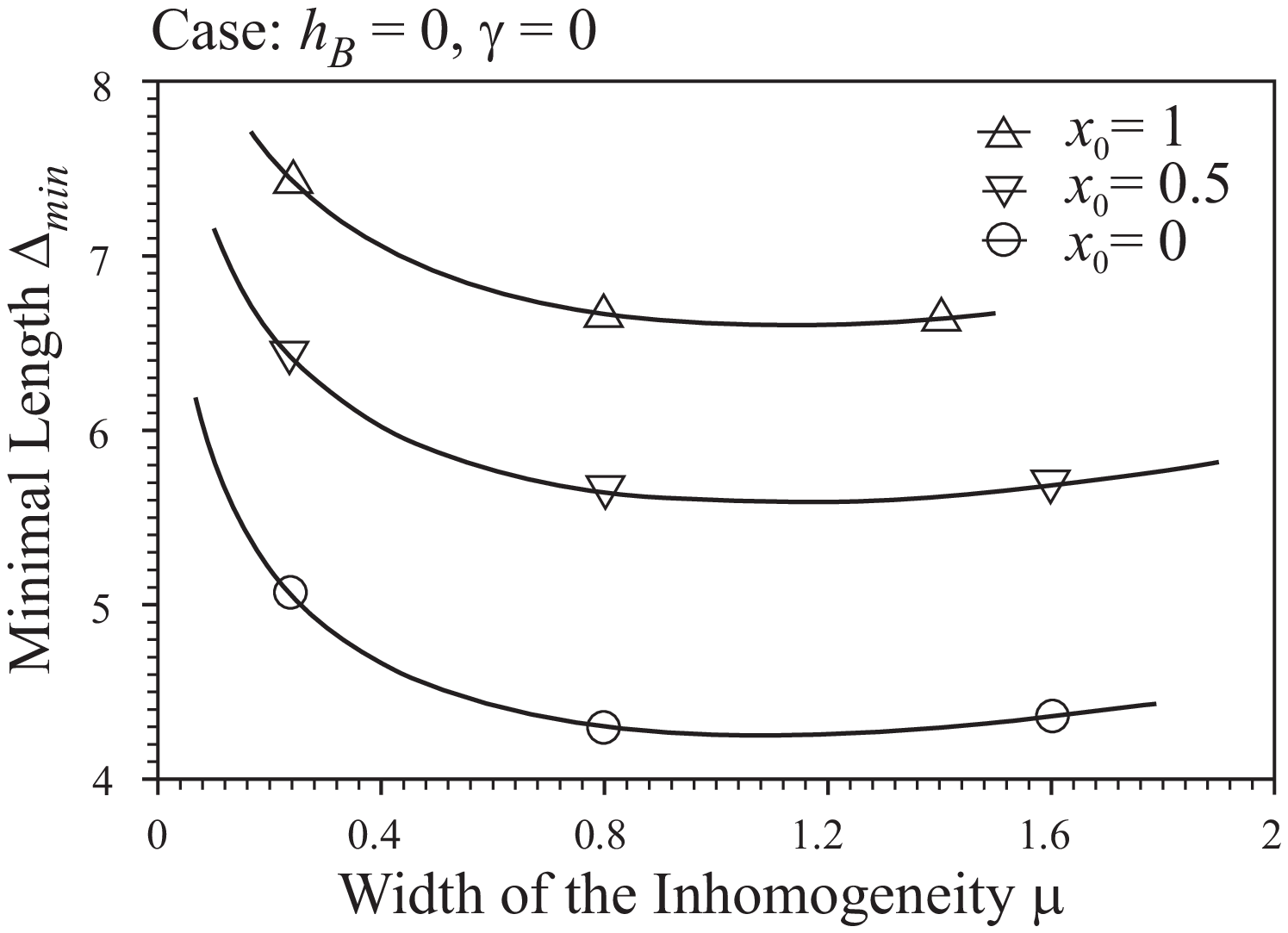,width=4.in}}
\centerline{\small Figure 11}\vspace{0.5cm}

At Fig.11 the numerically obtained relationship
$\Delta_{min}(\mu)$ is drawn, depending on the length of the
trapezium's base $\mu $ (width of the inhomogeneity) (see Fig.1).
The presented three curves corresponding to different locations
of the centre of inhomogeneity $x_0$ are qualitatively similar.
The most important their feature is the existence of minimums,
which are placed in the interval $\mu \in (0.9, 1.3)$. Therefore,
there exist such values of the quantity $\mu$, which provide a
minimal bifurcation length of a JJ from among all the minimal
lengths for given other parameters. Let us call them ``optimal"
lengths. For example, if $x_0 = 0$ (``symmetric" inhomogeneous
JJ), the optimal length $\Delta_{opt} \equiv \Delta_{min} \simeq
4.22$ of the JJ corresponds to $\mu \simeq 1.1$, whereas
$\Delta_{min} \simeq 4.48$ if the width of the inhomogeneity is
$\mu = 0.5$, and $\Delta_{min} \simeq 4.3$ for $\mu = 1.5$.

\vspace{0.3cm} \centerline{\psfig{figure=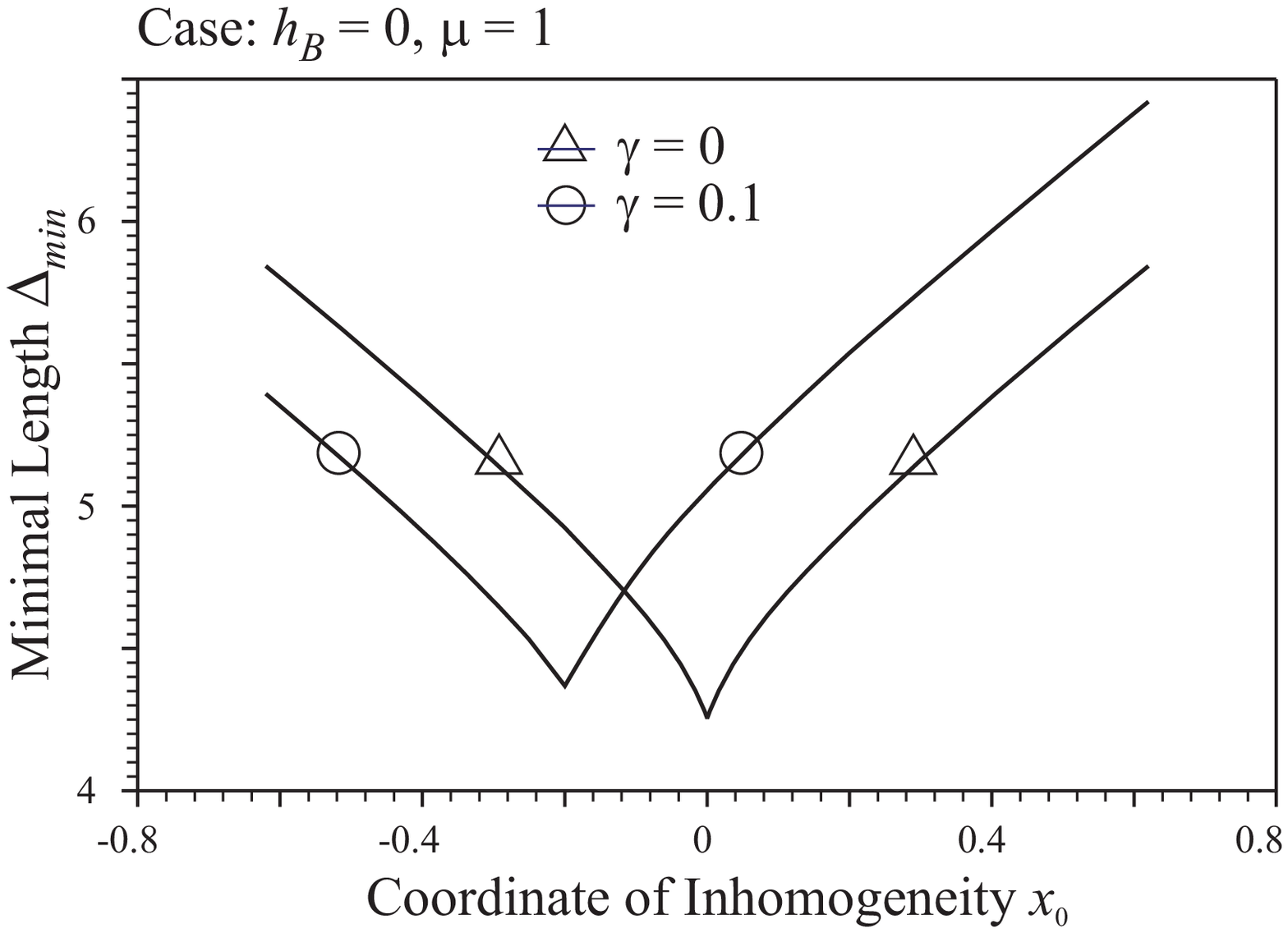,width=4.in}}
\centerline{\small Figure 12}\vspace{0.5cm}

The bifurcations of the bound states in the circular JJs with
respect to the location $x_0$ of the $\delta$-shaped
inhomogeneity are investigated numerically in \cite{vera}. Fig.12
shows the curves $\Delta_{min}(x_0)$ describing the behaviour of
the minimal length of the linear JJs. When the current $\gamma=0$
and the width of inhomogeneity $\mu=1$ is fixed, the curve is
symmetric and, apparently, the minimal value of $\Delta_{min}$ is
reached at the location $x_0=0$. When the current becomes
nontrivial ($\gamma = 0.1$), the minimum shifts into some other
point. For negative values of $\gamma$ the dependence is the same
but mirror reflected with respect to the line $x_0=0$.

\vspace{0.3cm} \centerline{\psfig{figure=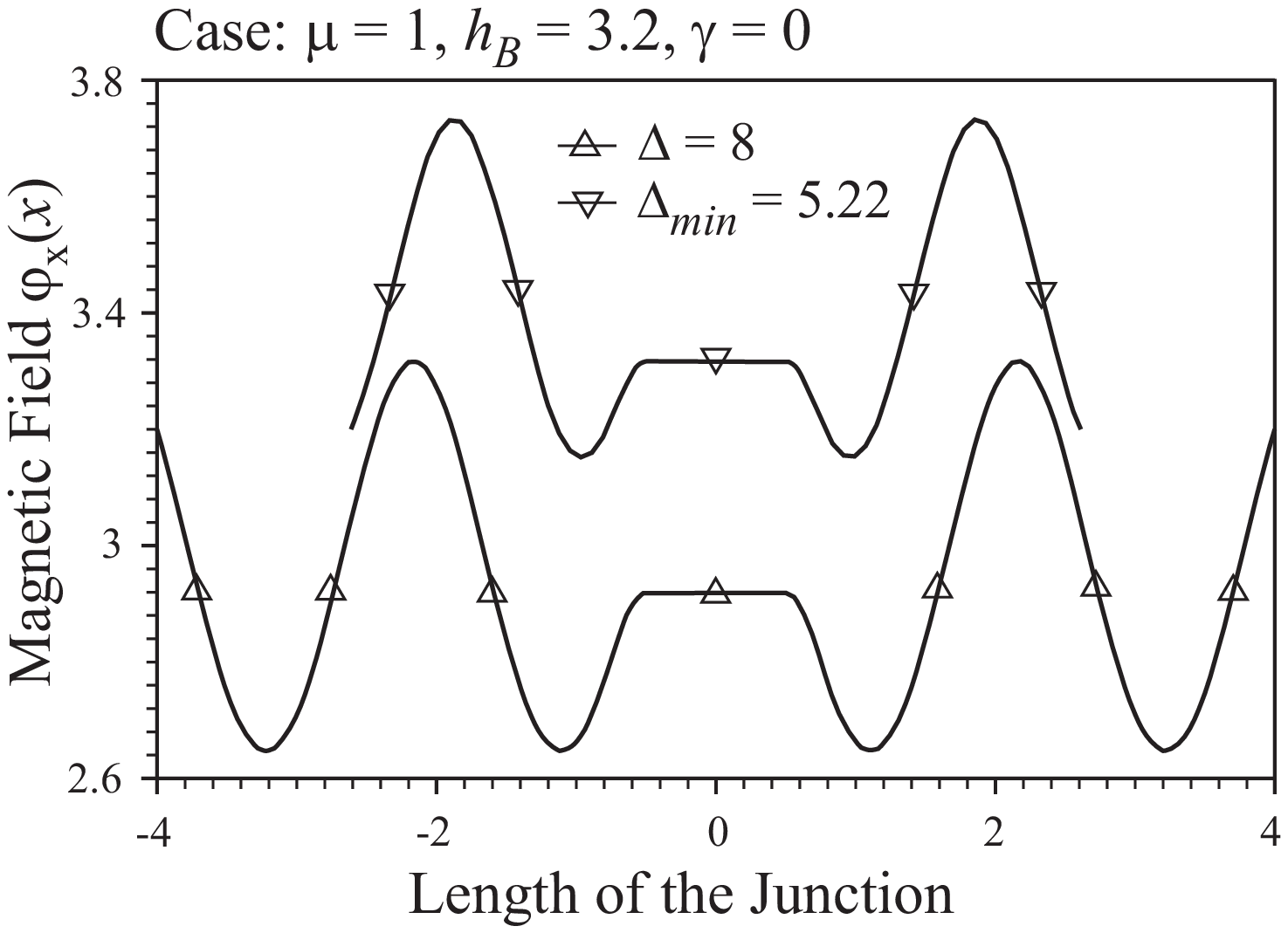,width=4.in}}
\centerline{\small Figure 13}\vspace{0.5cm}

Finally, the last Fig.13 provides as an illustration a
three-soliton bifurcation solution. Such kind of solutions can be
obtained when the boundary magnetic field $h_B$ is big enough.
Here, the concrete solution is derived for $h_B=3.2$. At the same
graph the initial distribution of the magnetic field
$\varphi_x(x)$ as solution of BVP (\ref{sg}), for $\Delta=8$,
$\gamma =0$, and $\mu = 1$ is included.

\medskip
{\bf Acknowledgements.} We thank Prof. I.V. Puzynin (JINR, Dubna,
Russia) for useful remarks. TLB remembers with gratitude the
elaborate and fruitful discussions with Prof. Yu.S. Gal'pern,
Prof. A.T. Filippov, and Prof. A.V. Ustinov in JINR, Dubna,
Russia.


\begin{thebibliography}{99}

\bibitem{barone}
Barone A and G.~Paterno G 1982 {\it Physics and Apllications of
the Jisephson Effect} (John Wiley \& Sons, Inc.)

\bibitem{and62} Anderson P W 1962 {\it Phys. Rev. Lett.} {\bf 9} 309

\bibitem{cfgsv00} Caputo J-G, Flytzanis N, Gaididei Y, Stefanakis N,
and Vavalis E (2000) {\it Superconductor Science and Technology}
{\bf 13} 423

\bibitem{mkls78} McLaughlin D W and Scott A C 1978 {\it Phys. Rev. A}
{\bf 18} 1652

\bibitem{galpfil} Gal'pern Yu S and Filippov A T 1984 {\it Sov.JETP} {\bf 86}(4)
1527 (in Russian)

\bibitem{alexboyad} Alexeeva N V and Boyadjiev T L 1997
{\it Bulg. J. of Physics}(1,2)

\bibitem{tedmi00}  Boyadjiev T L and Todorov M D 2000 {\it Mathematical Modeling
(Russian Academy of Sciences)} {\bf 12}(4) 61


\bibitem{vystavkin}  Vystavkin A N, Drachevski Yu F, Kosheletz V P, and
Serpuchenko I L 1988 {\it Fizika nizkih temperatur (Russian Phys.
of Low Temper.)} {\bf 14}(6) 646 (in Russian)

\bibitem{lmu94}  Larsen B H, Mygind J, and Ustinov A V 1994 {\it Phys. Lett. A}
{\bf 193} 359

\bibitem{mu94} Malomed B A, Ustinov A V 1994 {\it Phys. Rev. B}
{\bf 49}(18) 13024

\bibitem{boyad87}  Filippov A T, Gal'pern Yu S, Boyadjiev T L, and Puzynin I V
(1987) {\it Phys. Lett. A} {\bf 120}(1)

\bibitem{boyad88}  Boyadjiev T L, Pavlov D V, and Puzynin I V 1988
{\it Comm. of JINR, Dubna} P11-88-409 (in Russian)

\bibitem{paz}  Puzynin I V, Amirkhanov I V, Zemlyanaya E V, Pervushin V N,
Puzynina T P, Strizh T A, and Lakhno V D 1999 {\it Physics of
Elementary Particles and Atomic Nuclei}(JINR, Dubna) vol. 30, No
1, p.210 (in Russian), English translation:(American Institute of
Physics)p.97

\bibitem{levitan} Levitan B M and Sargsjan I S 1975
{\it Introduction to Spectral Theory}, Transl. Math.
Monographs(RI:AMS, Providence)

\bibitem{vera} Kaschieva V 1997 {\it Lecture Notes Comp. Sci.} {\bf 1196}
(Springer)pp 236--242


\end{thebibliography}
\end{document}